\documentstyle[12pt]{article}
\begin{document}        
\begin{titlepage}
\begin{flushright}
AMES-HET-98-02\\
IHEP-TH-98-01\\
TU-540\\
\vspace{.3cm}
hep-ph/9803334\\
\end{flushright}
\vspace{0.1in}
\begin{center}
{\Large Effective CP-violating operators of the tau lepton\\
         and some of their phenomenologies }
\vspace{.4in}

          Tao Huang$^d$, Jin Min Yang$^{a,b,c}$, Bing-Lin Young$^a$ 
          and Xinmin Zhang$^d$

\vspace{.4in}
\it

$^a$     Department of Physics and Astronomy\\ 
         and \\
$^b$     International Institute of Theoretical \& Applied Physics,\\
         Iowa State University, Ames, Iowa 50011, USA\\
$^c$     Department of Physics, Tohoku University, Sendai 980-77, Japan\\
$^d$     Institute of High Energy Physics, Academia Sinica, \\
         Beijing 100039, China
\rm
\end{center}
\vspace{3cm}

\begin{center} ABSTRACT\end{center}

The dimension-six CP-violating $SU_L(2)\times U_Y(1)$ invariant operators 
involving the tau lepton are studied.
The constraints from the available experimental data on tau dipole 
moments are 
derived. Under the current constraints, the induced  CP-violating 
effects could possibly be observed in
$\tau \rightarrow 3\pi \nu_{\tau}$ at the future tau-charm factory. 

\vfill
\end{titlepage}
\eject

\baselineskip=0.30in
\begin{center} {\Large 1. Introduction }\end{center}

The tau lepton is possibly a special probe of new physics in the leptonic
sector due to the fact that it is the only lepton
which is heavy enough to have hadronic decays and, as naively
expected, heavier fermions are more sensitive to the new physics related 
to mass generations. 
In searching for the possible new physics associated
with the tau lepton, CP-violation is a particularly interesting  probe.
In the standard model (SM), the existence of a phase in the 
Cabibbo-Kobayashi-Maskawa 
mixing matrix [1]  allows CP-violation in the quark sector but not 
in the lepton sector. The origin of CP-violation remains a mystery
after more than three decades of its discovery.
In models beyond the SM, additional CP-violation can appear 
rather naturally and non-CKM-type CP-violation is necessary in order 
to account for the observed value of baryon density to entropy ratio
[2]. Therefore, the detection of any non-CKM-type CP-violations,
such as CP-violating lepton interaction, will be an unequivocal signal
of new physics, and may help illuminating the origin of CP-violating and
alleviating the difficulty in baryogenesis.  

Experimental studies to date for CP violation
in tau processes have been in production and not in decay processes.
Since the bounds are relatively weak,
there are plenty rooms for CP-violating new physics to be discovered 
in the tau sector. With the tau-charm factory, 
which will operate at an $e^+e^-$ center-of-mass energy of around 4
GeV and a luminosity of $L=10^{33} $cm$^{-2}$ s$^{-1}$ with good 
$\pi/K$ separation, tau properties will be measured to a very high
precision. This will allow the tests of the SM and provide
some information about new physics. 

CP-violations in tau lepton decays have been investigated 
in Refs. [3-7].
Generally,  the possible CP-violating effects are expected  
to be larger and easier for detection in semileptonic tau decays 
than in production processes [8].
Analyses of new physics models yielding CP-violating effects in tau decays
have been given in Ref. [6], where the CP-violating effects of
multi-Higgs-doublet models are found to be possibly observable at the 
tau-charm factory.
More recently, a systematical analysis for the potential of the 
tau-charm factory in probing the CP violation of the tau sector 
has been given in [9].  

In this article, we give a model-independent study for
the possible CP-violating effects associated with tau lepton by
the use of the effective Lagrangian approach. 
The use of the effective Lagrangian approach in 
describing new physics is well motivated.
The fact that no direct signal of new
particles has been observed from collider experiments
and the impressive success of the SM requires that the new physics 
preserves the SM structure around the SM energy scale and only 
very delicately improves it [10].  
So it is likely that the only observable effects of new physics 
at energies not too far above the SM energy scale could be in the 
form of anomalous interactions which slightly affect the couplings 
of the SM particles.
 In this spirit, the residue effects of new physics can be expressed
as non-standard terms in an effective Lagrangian with a form like
\begin{equation}\label{eq1}
{\cal L}_{eff}={\cal L}_0+\frac{1}{\Lambda^2}\sum_i C_i O_i
                         +{\cal O} (\frac{1}{\Lambda^4}),
\end{equation}
where ${\cal L}_0$ is the SM Lagrangian, $\Lambda$ is the new physics
scale and $O_i$ are CP-conserving or CP-violating 
$SU_c(3)\times SU_L(2)\times U_Y(1)$ invariant
dimension-six operators,  and $C_i$ are
constants which represent the coupling strengths of $O_i$.
The expansion in Eq.(\ref{eq1}) was first discussed in Ref. [11]
and  further investigated in [12][13]][14][15]. In this article, we focus 
on the CP-violating operators involving the tau lepton.

In Sec.2 we list the possible dimension-six CP-violating
$SU_L(2)\times U_Y(1)$ invariant operators involving the tau lepton and
give their expressions  after electroweak gauge symmetry breaking.
In Sec.3 we give the induced CP-violating
effective couplings  $W\nu \tau $, $Z\tau \tau$ and
$\gamma \tau \tau$  and classify the operators according to
the interaction vertices.
In Sec.4 we derive the bounds for the coupling strength from
the available experimental data in tau dipole moments.
In Sec.5 we evaluate the possibility of observing 
the CP-violating effects of these operators 
in $\tau \rightarrow 3\pi \nu_{\tau}$ at a future tau-charm
factory.  And finally in Sec.6 we present the summary.
\vspace{.5cm}

\begin{center} {\Large 2. CP-violating operators involving the tau lepton}\end{center}
\vspace{.5cm}

Here we assume that the new physics in the 
lepton sector resides in the interaction of third family 
to gauge bosons or Higgs boson. 
Therefore, the operators we are interested in  are those containing 
third-family leptons coupling to gauge or Higgs bosons.

To restrict ourselves to the lowest order, we consider only tree diagrams
and to the order of $1/\Lambda^2$. Therefore, only one
vertex in a given diagram can contain anomalous couplings. Under these 
conditions, operators which are related by the field equations
are not independent. As discussed in Ref.[14], to which we refer 
for the detail, the fermion and the Higgs boson equations of motion 
can be used but the equations of motion of the gauge bosons can not when 
writing down the operators in Eq.(\ref{eq1}).
Also, we assume all the operators $O_i$ to be Hermitian. 
Because of our assumption 
that the available energies are below the unitarity cuts of new-physics 
particles, no imaginary part can be generated by the new physics effect. 
Therefore the coefficients $C_i$ in Eq.(\ref{eq1}) are real. 

The expressions of the CP-violating operators involving the third family 
leptons are parallel to their corresponding ones involving the third family 
quarks given in Ref.[15], but
the number of independent operators is much less due to the absence of 
right-handed neutrino and the strong interactions. 
We follow the standard notation:
$L$ denotes the third family left-handed doublet leptons, $\Phi$ is 
the Higgs doublet, $W_{\mu\nu}$ and $B_{\mu\nu}$ are the SU(2) and 
U(1) gauge boson field tensors in the appropriate matrix forms, and
$D_\mu$ denotes the appropriate covariant derivatives.  For more
details of the notation we refer to Ref.[14].
The possible operators are given by:
\begin{eqnarray}
O_{LW}&=&i\left [\bar L \gamma^{\mu}\tau^I D^{\nu}L
         -\overline{D^{\nu}L} \gamma^{\mu}\tau^I L\right ] W^I_{\mu\nu},\\
O_{LB}&=&i\left [\bar L \gamma^{\mu} D^{\nu}L
         -\overline{D^{\nu}L} \gamma^{\mu} L\right ] B_{\mu\nu},\\
O_{\tau B}&=&i\left [\bar \tau_R \gamma^{\mu} D^{\nu}\tau_R
         -\overline{D^{\nu}\tau_R} \gamma^{\mu}\tau_R\right ] B_{\mu\nu},\\
O_{\Phi L}^{(1)}&=&\left [\Phi^{\dagger}D_{\mu}\Phi
                   +(D_{\mu}\Phi)^{\dagger}\Phi\right ] \bar L \gamma^{\mu}L,\\
O_{\Phi L}^{(3)}&=&\left [\Phi^{\dagger}\tau^I D_{\mu}\Phi
                   +(D_{\mu}\Phi)^{\dagger}\tau^I\Phi\right ]
                   \bar L \gamma^{\mu}\tau^I L,\\
O_{\Phi \tau}&=&\left [\Phi^{\dagger}D_{\mu}\Phi
             +(D_{\mu}\Phi)^{\dagger}\Phi\right ]
             \bar \tau_R \gamma^{\mu}\tau_R,\\
O_{\tau 1}&=&i(\Phi^{\dagger}\Phi-\frac{v^2}{2})\left [\bar L \tau_R\Phi
         -\Phi^{\dagger}\bar \tau_R L\right ],\\
O_{D\tau}&=&i\left [(\bar L D_{\mu} \tau_R) D^{\mu}\Phi
         -(D^{\mu}\Phi)^{\dagger}(\overline{D_{\mu}\tau_R}L)\right ],\\
O_{\tau W\Phi}&=&i\left [(\bar L \sigma^{\mu\nu}\tau^I \tau_R) \Phi
         -\Phi^{\dagger}(\bar \tau_R \sigma^{\mu\nu}\tau^I L)\right ]
          W^I_{\mu\nu},\\
O_{\tau B\Phi}&=&i\left [(\bar L \sigma^{\mu\nu} \tau_R) \Phi
         -\Phi^{\dagger}(\bar \tau_R \sigma^{\mu\nu} L)\right ]
          B_{\mu\nu}.
\end{eqnarray}

The expressions of these CP-violating operators 
after electroweak symmetry breaking  in the unitary gauge
are given by
\begin{eqnarray}\label{eq12}
O_{LW}&=&\frac{i}{2}W^3_{\mu\nu}
\left [\bar \nu_{\tau}\gamma^{\mu}P_L\partial^{\nu}\nu_{\tau}
-\partial^{\nu}\bar \nu_{\tau}\gamma^{\mu}P_L\nu_{\tau}
-\bar \tau \gamma^{\mu}P_L \partial^{\nu}\tau
+\partial^{\nu}\bar \tau\gamma^{\mu}P_L\tau\right ]     \nonumber\\
& & +\frac{i}{\sqrt 2}
\left [W^+_{\mu\nu}(\bar \nu_{\tau}\gamma^{\mu}P_L\partial^{\nu}\tau
-\partial^{\nu}\bar \nu_{\tau}\gamma^{\mu}P_L\tau)
+W^-_{\mu\nu}(\bar \tau\gamma^{\mu}P_L\partial^{\nu}\nu_{\tau}
-\partial^{\nu}\bar \tau\gamma^{\mu}P_L\nu_{\tau})\right ]\nonumber\\
& & +g_2\bar L \gamma^{\mu} \left [W_{\mu},W_{\nu}\right ]\partial^{\nu}L
  -g_2\partial^{\nu}\bar L \gamma^{\mu} 
  \left [W_{\mu},W_{\nu}\right ]L\nonumber\\
& &+\frac{1}{2}g_2 (\vec W_{\mu\nu}\cdot \vec W^{\nu})
   \bar L \gamma^{\mu} L
  -g_1 B^{\nu} \bar L \gamma^{\mu}W_{\mu\nu} L,\\
\label{eq13}
O_{LB}&=&iB_{\mu\nu}\left [\bar L\gamma^{\mu} \partial^{\nu}L
   -\partial^{\nu} \bar L\gamma^{\mu}L
   -2i\bar L\gamma^{\mu}(g_2 W^{\nu}
   -\frac{1}{2}g_1B^{\nu})L\right ],\\
\label{eq14}
O_{\tau B}&=&i\left [\bar \tau_R\gamma^{\mu} \partial^{\nu}\tau_R
         -\partial^{\nu} \bar \tau_R\gamma^{\mu}\tau_R\right ]B_{\mu\nu}
      -2g_1\bar \tau_R\gamma^{\mu}\tau_R B_{\mu\nu}B^{\nu},\\
O_{\Phi L}^{(1)}&=&(H+v)\partial_{\mu}H 
\left [\bar \nu_{\tau}\gamma^{\mu}P_L\nu_{\tau}
          +\bar \tau\gamma^{\mu}P_L\tau\right ], \\
\label{eq16}
O_{\Phi L}^{(3)}&=&-O_{\Phi L}^{(1)}+2(H+v)\partial_{\mu}H 
        \bar \tau\gamma^{\mu}P_L\tau
     -\frac{ig_2}{\sqrt 2}(H+v)^2(W^+_{\mu}\bar \nu_{\tau}\gamma^{\mu}P_L\tau
                -W^-_{\mu}\bar \tau\gamma^{\mu}P_L\nu_{\tau}),\\
O_{\Phi \tau}&=&(H+v)\partial_{\mu}H \bar \tau \gamma^{\mu}P_R\tau,\\
O_{\tau 1}&=&\frac{1}{2\sqrt 2}H(H+v)(H+2v)\bar \tau i\gamma_5 \tau,\\
O_{D\tau}&=&i\frac{1}{2\sqrt 2}\partial^{\mu}H \left [
   \bar \tau\partial_{\mu}\tau-(\partial_{\mu}\bar \tau)\tau
   +\partial_{\mu}(\bar \tau\gamma_5 \tau)
   +i2g_1B_{\mu}\bar \tau\tau\right ]\nonumber\\
& & +\frac{1}{2\sqrt 2}\frac{m_Z}{v} (H+v)Z^{\mu}
\left [\partial_{\mu}(\bar \tau\tau)
    +\bar \tau\gamma_5\partial_{\mu}\tau-(\partial_{\mu}\bar \tau)\gamma_5 \tau
    +2g_1B_{\mu}(\bar \tau i\gamma_5 \tau)\right ]\nonumber\\
& & +\frac{g_2}{2}(H+v)
\left [W_{\mu}^+ (\bar \nu_{\tau}P_R \partial_{\mu}\tau
    +ig_1 B_{\mu} \bar \nu_{\tau}P_R  \tau)
+W_{\mu}^- ( \partial_{\mu}\bar \tau P_L \nu_{\tau}-
                 ig_1 B_{\mu} \bar \tau P_L \nu_{\tau})\right ],\\
O_{\tau W\Phi}&=&i\frac{1}{2}(H+v)
\left [W^+_{\mu\nu}(\bar \nu_{\tau} \sigma^{\mu\nu}P_R \tau)
    -W^-_{\mu\nu}(\bar \tau\sigma^{\mu\nu}P_L \nu_{\tau})
    -\frac{1}{\sqrt 2}W^3_{\mu\nu}
    (\bar \tau \sigma^{\mu\nu}\gamma_5 \tau)\right.\nonumber\\
& & +ig_2(W^+_{\mu}W^3_{\nu}-W^3_{\mu}W^+_{\nu})
(\bar \nu_{\tau} \sigma^{\mu\nu} P_R \tau)
    +ig_2(W^-_{\mu}W^3_{\nu}-W^3_{\mu}W^-_{\nu})
(\bar \tau \sigma^{\mu\nu}P_L \nu_{\tau})
\nonumber\\
& &\left. +i\frac{g_2}{\sqrt 2} (W^+_{\mu}W^-_{\nu}-W^-_{\mu}W^+_{\nu})
      (\bar \tau \sigma^{\mu\nu}\gamma_5 \tau)\right ],\\
\label{eq21}    
O_{\tau B\Phi}&=&\frac{i}{\sqrt 2}(H+v) B_{\mu\nu}
            (\bar \tau \sigma^{\mu\nu}\gamma_5 \tau),
\end{eqnarray}
where we use the convention $Z_{\mu}=-\cos\theta_W W^3_{\mu}
+\sin\theta_W B_{\mu}$ and $A_{\mu}=\sin\theta_W W^3_{\mu}
+\cos\theta_W B_{\mu}$. 
Note that most of the above operators clearly show the $U_{em}(1)$ 
gauge invariance,
while some of them do not manifest this invariance straight forwardly. 
We have checked that all the operators listed above give indeed a 
$U_{em}(1)$ gauge invariant expression.
\vspace{1cm}

\begin{center} {\Large 3.~Effective vertices for the gauge couplings of tau}
                \end{center}


The possibilities of contributions of the dimension-six CP-violating operators 
to some three-particle couplings are shown in Table 1. 
According to their contribution to the three-particle vertices of
charged and neutral current, we classify the operators as: \\
 Class A:  $O_{\Phi L}^{ (3)}$, contributing to charged current.\\
 Class B:  $O_{LW}$, $O_{D\tau}$ and $O_{\tau W\Phi}$, contributing to both 
            charged and neutral currents.\\
 Class C:  $O_{LB}$, $O_{\tau B}$ and $O_{\tau B\Phi}$, contributing  to
                                               neutral currents.\\
 Class D:  $O_{\Phi L}^{ (1)}$, $O_{\Phi \tau}$ and $O_{\tau 1}$, 
                no contribution to charged and neutral currents.

Since Class D operators contribute only to the $H\tau\tau$ coupling, they 
may not be probed at future colliders. We will not consider
these operators here further. 
Both Class B and Class C operators affect neutral currents of the tau,
and, as our analysis show, they will be strongly constrained by
LEP data for the dipole moments of the tau. 
Since the constraints for the charged current of the tau from 
its leptonic decays are much weaker (see below), Class A operator
will not be strongly constrained and, as a result, will provide the best
possibility for the observation of CP-violating
effects in hadronic tau decays. 
  
Collecting all the relevant terms we get the effective CP-violating 
couplings,
\begin{eqnarray}
{\cal L}_{W\nu\tau}&=&-i\frac{C_{\Phi L}^{(3)}}{\Lambda^2}\frac {g_2}{\sqrt 2}
      v^2 W^+_{\mu}(\bar \nu_{\tau}\gamma^{\mu}P_L \tau)
     -i\frac{C_{D\tau}}{\Lambda^2}\frac{v}{\sqrt 2}\frac{g_2}{\sqrt 2} 
      W_{\mu}^+ \bar \nu_{\tau} P_R (i\partial^{\mu}\tau)\nonumber\\
& & +i\frac{C_{\tau W\Phi}}{\Lambda^2}\frac{v}{2} 
      W^+_{\mu\nu}(\bar \nu_{\tau} \sigma^{\mu\nu}P_R \tau)
   +i\frac{C_{LW}}{\Lambda^2} \frac{1}{\sqrt 2}
        W^+_{\mu\nu}[\bar \nu_{\tau}\gamma^{\mu}P_L(\partial^{\nu}\tau)
                     -(\partial^{\nu}\bar \nu_{\tau})\gamma^{\mu}P_L \tau],\\
{\cal L}_{Z\tau\tau}&=&
i(\frac{C_{\tau W\Phi}}{\Lambda^2}\frac{c_Wv}{2\sqrt 2}
+\frac{C_{\tau B\Phi}}{\Lambda^2}\frac{v}{\sqrt 2}s_W) 
   Z_{\mu\nu}(\bar \tau \sigma^{\mu\nu}\gamma_5 \tau)\nonumber\\
& & +i(\frac{C_{LW}}{\Lambda^2}\frac{c_W}{2}
                  +\frac{C_{LB}}{\Lambda^2}s_W) 
   Z_{\mu\nu}(\bar \tau \gamma^{\mu}P_L \partial^{\nu}\tau
                      -\partial^{\nu}\bar \tau \gamma^{\mu}P_L \tau)\nonumber\\
& & +i\frac{C_{\tau B}}{\Lambda^2}s_W
    Z_{\mu\nu}(\bar \tau \gamma^{\mu}P_R \partial^{\nu}\tau
                -\partial^{\nu} \bar \tau\gamma^{\mu}P_R \tau)\nonumber\\
& &-i\frac{m_Z}{2\sqrt 2} \frac{C_{D\tau}}{\Lambda^2} 
 Z^{\mu}\left [i(\bar \tau\gamma_5\partial_{\mu}\tau
  -\partial_{\mu}\bar \tau\gamma_5 \tau)
   +i\partial_{\mu}(\bar \tau\tau )
      \right ],\\
{\cal L}_{\gamma \tau\tau}&=&
 i (\frac{C_{LB}}{\Lambda^2}c_W-\frac{C_{LW}}{\Lambda^2}\frac{s_W}{2})
  A_{\mu\nu}(\bar \tau \gamma^{\mu}P_L \partial^{\nu}\tau
           -\partial^{\nu}\bar \tau \gamma^{\mu}P_L \tau)\nonumber\\
& & +i\frac{C_{\tau B}}{\Lambda^2}c_W
   A_{\mu\nu}(\bar \tau \gamma^{\mu}P_R \partial^{\nu}\tau
         -\partial^{\nu} \bar \tau\gamma^{\mu}P_R \tau)\nonumber\\
& &+i(\frac{C_{\tau B\Phi}}{\Lambda^2}c_W
     -\frac{C_{\tau W\Phi}}{\Lambda^2}\frac{s_W}{2})\frac{v}{\sqrt 2}
   A_{\mu\nu}(\bar \tau \sigma^{\mu\nu}\gamma_5 \tau),
\end{eqnarray}
where  $s_W\equiv \sin\theta_W$, $c_W\equiv \cos\theta_W$ and
$P_{L,R}\equiv(1\mp \gamma_5)/2$.
\vspace{1cm}

\begin{center} {\Large 4. Current constraints from 
                          experimental data} \end{center}

\begin{center} 
{\large 4.1 Constraints from the measurement of the dipole moments}
  \end{center}

Including both the SM couplings and CP-violating new physics effects,  
we can write the $V\tau\tau$ $(V=Z,\gamma)$ vertices, with
both taus being on-shell, as
\begin{equation}\label{ver}
 \Gamma^{\mu}_{V\tau\tau}=ie\left [\gamma_{\mu}A_V
-\gamma_{\mu}\gamma_5 B_V+\frac{C_V}{2m_{\tau}}k_{\nu}\sigma^{\mu\nu}\gamma_5
\right ],
\end{equation}
where $k$ is the momentum of the vector boson.
We have neglected  the scalar and pseudo-scalar couplings, 
$k_{\mu}$ and $k_{\mu}\gamma_5$, since 
these terms give contributions proportional
to the electron mass in $e^+e^-\rightarrow \tau^-\tau^+$. 
We note that some of these neglected terms are
needed to maintain the electromagnetic gauge invariance for the axial
vector couplings in Eq.(\ref{ver}).
$A_V$ and $B_V$ are the SM couplings. At the tree level, they are given by
$A_{Z,\gamma}=\frac{-1+4s_W^2}{4s_Wc_W}, -1$ and
$B_{Z,\gamma}=\frac{-1}{4s_Wc_W}, 0$.
$C_V$ arises from new physics given by 
\begin{eqnarray}
C_Z&=&-\frac{C_{LW}}{\Lambda^2}\frac{m_{\tau}^2}{e}c_W
             +\frac{C_{D\tau}}{\Lambda^2}\frac{m_{\tau}m_Z}{\sqrt 2 e}
+\frac{C_{\tau W\Phi}}{\Lambda^2}\frac{vm_{\tau}}{e}\sqrt 2 c_W\nonumber\\
& &  -\frac{C_{LB}-C_{\tau B}}{\Lambda^2}\frac{m_{\tau}^2}{e}2 s_W
    +\frac{C_{\tau B\Phi}}{\Lambda^2}\frac{vm_{\tau}}{e}2\sqrt 2 s_W ,\\
C_{\gamma}&=&-\frac{C_{LW}}{\Lambda^2}\frac{m_{\tau}^2}{e}s_W
+\frac{C_{\tau W\Phi}}{\Lambda^2}\frac{vm_{\tau}}{e}
                    \sqrt 2 s_W\nonumber\\
& & +\frac{C_{LB}-C_{\tau B}}{\Lambda^2}\frac{m_{\tau}^2}{e}2 c_W
-\frac{C_{\tau B\Phi}}{\Lambda^2}\frac{vm_{\tau}}{e}2\sqrt 2 c_W.
\end{eqnarray}
The electric and weak dipole moments are obtained by
\begin{equation}
d^{\gamma, Z}_{\tau}=\frac{e}{2m_{\tau}}C_{\gamma, Z}
                     =0.55\times 10^{-14} C_{\gamma, Z} ~~({\rm e~cm}).
\end{equation}

The CP-violation introduced by the dipole moments can be searched 
in $Z\rightarrow \tau^+\tau^-$. The dipole moments can be determined 
from the tau spin which can be measured from the tau decay products.
No evidence of CP-violation has been observed in 
 $Z\rightarrow \tau^+\tau^-$ so far, which set strong limits on the dipole
moments.

The limit on the weak dipole moment of tau lepton obtained
at LEP is [16] 
\begin{eqnarray} \label{weak}
\vert {\rm Re}~d^{Z}_{\tau} \vert &\le & 3.6 \times 10^{-18}~~ {\rm e~cm}~~~(95\% ~C.L.).
\end{eqnarray}
Assuming the simple situation that cancellation among different 
operators does not take place, we get the bounds on the coupling
strength 
\begin{eqnarray}\label{neu1}
\left \vert \frac{C_{LW}}{\Lambda^2} \right \vert &<&6.8\times 10^{-5} ~{\rm GeV}^{-2},\\
\left \vert \frac{C_{D\tau}}{\Lambda^2} \right \vert &<&1.7\times 10^{-6} ~{\rm GeV}^{-2},\\
\left \vert \frac{C_{\tau W\Phi}}{\Lambda^2} \right \vert &<&3.5\times 10^{-7} ~{\rm GeV}^{-2},\\
\left \vert \frac{C_{LB}}{\Lambda^2} \right\vert, 
\left \vert \frac{C_{\tau B}}{\Lambda^2} \right\vert 
            &<& 6.3 \times 10^{-5} ~{\rm GeV}^{-2},\\  \label{neu5}
\left \vert \frac{C_{\tau B\Phi}}{\Lambda^2} \right \vert &<&3.2\times 10^{-7} ~{\rm GeV}^{-2}.
\end{eqnarray}

Compared with the constraints on the weak dipole moment,  those 
on the electric dipole moment of tau lepton are weaker. 
The strongest constraint on $d^{\gamma}_{\tau}$ has been derived from
the $Z\rightarrow \tau^+\tau^-$ decay width, which is given by [17]
\begin{eqnarray}
 \vert d^{\gamma}_{\tau} \vert  < 2.7 \times 10^{-17}~~ {\rm e~cm} ~~~(95\%~C.L.),
\end{eqnarray}
which yield the following bounds on the coupling strength 
under the assumption that cancellation among different 
operators does not take place
\begin{eqnarray}
\left \vert \frac{C_{LW}}{\Lambda^2} \right\vert &<&9.4\times 10^{-4} ~{\rm GeV}^{-2},\\
\left \vert \frac{C_{\tau W\Phi}}{\Lambda^2} \right\vert &<&4.9\times 10^{-6} ~{\rm GeV}^{-2},\\
\left \vert \frac{C_{LB}}{\Lambda^2} \right\vert, 
\left\vert \frac{C_{\tau B}}{\Lambda^2} \right\vert 
            &<& 2.6 \times 10^{-4} ~{\rm GeV}^{-2},\\  
\left\vert \frac{C_{\tau B\Phi}}{\Lambda^2} \right\vert &<&1.3\times 10^{-6} ~{\rm GeV}^{-2}.
\end{eqnarray}
\vspace{.5cm}

\begin{center} 
{\Large 4.2 Constraints from the measurement of leptonic decays}
  \end{center}

The CP-violating contribution to the 
$W \nu \tau$ vertex in the tau decay can be written in the momentum space as
\begin{eqnarray}
 {\cal L}_{W\nu \tau}&=&\frac{g_2}{\sqrt 2} W_{\mu}^+\bar \nu_{\tau}
 \left [ \gamma_{\mu} P_L (1+ia)+k_{\mu}P_R \frac{1}{2m_{\tau}}(ib)
+\frac{i}{2m_{\tau}}k^{\nu}\sigma_{\mu\nu}P_R (ic) \right ] \tau ,
\end{eqnarray}
where the form factors are given by
\begin{eqnarray}
a&=&-\frac{C^{(3)}_{\Phi L}}{\Lambda^2}v^2
  -\frac{C_{D\tau}}{\Lambda^2} \frac{vm_{\tau}}{2\sqrt 2},\\
b&=&-\frac{C_{D\tau}}{\Lambda^2} \frac{vm_{\tau}}{\sqrt 2},\\
c&=&\frac{C_{LW}}{\Lambda^2} \frac{2m_{\tau}^2}{g_2}
-\frac{C_{D\tau}}{\Lambda^2} \frac{vm_{\tau}}{\sqrt 2}
-\frac{C_{\tau W\Phi}}{\Lambda^2} \frac{2\sqrt 2vm_{\tau}}{g_2}.
\end{eqnarray}
The $k_{\mu}$ term is negligible in leptonic tau decays.
The theoretic prediction for branching fractions of the 
decay $\tau^-\rightarrow l^-\bar \nu_l\nu_{\tau}$ ($l=e^-,\mu^-$)
are given by [18][19][20]
\begin{eqnarray}
B_l&=&\frac{G^2_F m_{\tau}^5}{192 \pi^3}\tau_{\tau} (1-8x-12x^2\ln x+8x^3
      -x^4) \nonumber\\
   & & \times \left [ \left ( 1-\frac{\alpha(m_{\tau})}{2\pi}
        (\pi^2-\frac{25}{4})\right ) 
  \left ( 1+\frac{3}{5}\frac{m_{\tau}^2}{m_W^2}-2\frac{m_l^2}{m_W^2}\right )
  \right ] ( 1+\Delta_l ),
\end{eqnarray}
where $\tau_{\tau}$ is the tau lifetime, $x=m_l^2/m_{\tau}^2$
and $\Delta_l=\tilde \kappa^2/10$ with $\tilde \kappa=\sqrt {c^2+10a^2}$.
The world average values for $B_l$ [21] constraint 
$\vert \tilde \kappa\vert <0.26 $ at 95\% CL [20]. 
Again assuming the simple situation that cancellation among different 
operators does not take place, this yields 
the upper bounds on coupling strengths of the operators
\begin{eqnarray} \label{phiq}
\left\vert \frac{ C^{(3)}_{\Phi L}}{\Lambda^2}\right \vert 
&<& 1.4 \times 10^{-6}~{\rm GeV}^{-2},\\
\left\vert \frac{C_{LW}}{\Lambda^2}\right \vert &<&2.4\times 10^{-2} ~{\rm GeV}^{-2},\\
\left\vert \frac{C_{D\tau}}{\Lambda^2} \right\vert &<&4.4\times 10^{-4} ~{\rm GeV}^{-2},\\
\left\vert \frac{C_{\tau W\Phi}}{\Lambda^2} \right\vert &<&1.2\times 10^{-4} ~{\rm GeV}^{-2}.
\end{eqnarray}
These bounds on $C_{LW}$, $C_{D\tau}$ and $C_{\tau W\Phi}$ derived 
from charged current are much weaker than those derived from neutral currents 
given in Eqs.(\ref{neu1}-\ref{neu5}). 

We summarize that the strongest bounds presently available on the seven operators 
in classes A, B, and C are given in Eqs. (\ref{neu1})-(\ref{neu5}) and (\ref{phiq}).
\vspace{.5cm} 

\begin{center} {\Large 5. CP-violating effects 
         in $\tau \rightarrow 3\pi \nu_{\tau}$ } \end{center}

As pointed out in Sec. 1, there are various methods in searching for the 
CP-violations of the tau lepton. Here we focus on the three-charged-pion decay  
$\tau^{\pm} \rightarrow \pi^{\pm} +\pi^{\mp}+\pi^{\pm}+\nu_{\tau}$,
which has been argued to be a promising process for detecting CP 
violation [6][7].
This decay is dominated by the contributions of two overlapping resonances,
$a_1$(1260) and $\pi'$(1300), and has been extensively studied [22][23].
In our analyses we follow Ref. [23] for the phenomenological parameterization 
of the form factors. 

Including the contributions of possible new physics,
the matrix element for the parton-level process 
$\tau^-(p, \sigma) \rightarrow \bar u +d+\nu_{\tau}(k,-)$,
where the momenta and helicities for $\tau$ and $\nu_{\tau}$ are indicated,
is given by
\begin{eqnarray}
M&=&\sqrt 2 G_F \left [ 
(1+\chi) \bar u(k,-)\gamma_{\mu}P_L u(p,\sigma)
 ~\bar d\gamma^{\mu}(1-\gamma_5) u \right. \nonumber \\
 & & \hspace{1cm} +\eta \bar u(k,-)P_R u(p,\sigma)~\bar d (1+\gamma_5) u\nonumber \\
 & & \hspace{1cm} \left. +\zeta\bar u(k,-)\frac{p_{\mu}}{m_{\tau}} P_R u(p,\sigma)
 ~\bar d\gamma^{\mu}(1-\gamma_5) u \right ].
\end{eqnarray}
The form factors $\chi$, $\eta$ and $\zeta$, which are from new physics,
are given by
\begin{eqnarray}
\chi    &=& i (T_1+T_2+T_3), \\
\eta    &=& -i \frac{m_u+m_d}{m_{\tau}} (T_2+T_3), \\
\zeta &=& -i2(T_2+T_3+T_4), 
\end{eqnarray}
where 
\begin{eqnarray}
T_1 &=& -\frac{C^{(3)}_{\Phi L}} {\Lambda^2} v^2 , \\
T_2 &=&  \frac{C_{\tau W \Phi}} {\Lambda^2} \frac{\sqrt 2 v m_{\tau}}{g_2}, \\
T_3 &=& -\frac{C_{LW}} {\Lambda^2} \frac{m_{\tau}^2}{g_2}, \\
T_4 &=& \frac{C_{D\tau}} {\Lambda^2} \frac{v m_{\tau}}{2\sqrt 2}
\end{eqnarray}
Here $m_u$ and $m_d$ are the current
masses of the $u$ and $d$ quarks. 
Under the constraints derived in Sec. 4, we have, 
\begin{eqnarray}
\vert T_1 \vert &<& 8.47\times 10^{-2}, \\
\vert T_2 \vert &<& 3.65\times 10^{-4}, \\
\vert T_3 \vert &<& 3.67\times 10^{-4}, \\
\vert T_4 \vert &<& 2.65\times 10^{-4}. 
\end{eqnarray}
Here we see that the bound on term $T_1$ is much weaker than those on 
the other terms.   
In the following we only present the detailed analyses for term $T_1$, i.e. 
the effects of operator $O^{(3)}_{\Phi L}$.
Then the matrix element for the decay $\tau^-\rightarrow (3\pi)^-\nu_\tau$
can be written in the form:
\begin{eqnarray}
M=\sqrt{2}G_F (1+\chi)\bar{u}(k,-)\gamma^\mu P_L u(p,\sigma)J_\mu,
\end{eqnarray}
where $J_\mu$ is the vector hadronic matrix element given by [23]
\begin{eqnarray}
J_{\mu}&= &<(3\pi)^-|\bar{d}\gamma_\mu (1-\gamma_5) u|0>, \nonumber\\
       &= &N\left \{\frac{2\sqrt{2}}{3}T^{\mu\nu}
   \left [ (q_2-q_3)_\nu F_1(q^2,s_1)
   +(q_1-q_3)_\nu F_2(q^2,s_2)\right ]\right. \nonumber\\
   & &\left. \hskip 1.3cm +q^\mu C_{\pi^\prime}\left [ s_1(s_2-s_3)F_3(q^2,s_1)
      +s_2(s_1-s_3)F_4(q^2,s_2)\right ] \right \}.
\end{eqnarray}
Here $q_1$ and $q_2$ are the momenta of two identical $\pi^-$'s,
$q_3$ is the momentum of the $\pi^+$, $q$ is the momentum
of the $(3\pi)^-$ system, and $T^{\mu\nu}=g^{\mu\nu}-q^\mu q^\nu/q^2$.
$F_i$ are the form factors [23], with $F_3$ and  $F_4$ being related by Bose 
symmetry under $q_1\leftrightarrow q_2$.  
The kinematic invariants $s_i$ are defined by
\begin{eqnarray}
s_1=(q_2+q_3)^2,~~s_2=(q_3+q_1)^2,~~ s_3=(q_1+q_2)^2.
\end{eqnarray}
The constants $N$ and $C_{\pi^\prime}$ are obtained by [23]
\begin{eqnarray}
N&=&\frac{\cos\theta_C}{f_\pi}, \\
\label{Cpi}
C_{\pi^\prime}&=&\frac{g_{\pi^\prime\rho\pi}g_{\rho\pi\pi}
                       f_{\pi^\prime}f_\pi}{m^4_\rho m^2_{\pi^\prime}},
\end{eqnarray}
where $\theta_C$ is the Cabibbo angle,
and $g_{\pi'\rho\pi}$ and  $g_{\rho\pi\pi}$ are the strong coupling constants
of $\pi'$-$\rho$-$\pi$ and $\rho-\pi-\pi$, respectively.
The values of the relevant parameters are given as [23]
\begin{eqnarray}
&&\cos\theta_C=0.973,         ~~m_\rho=0.773\ \ {\rm GeV},\nonumber\\
&& g_{\pi^\prime\rho\pi}=5.8, ~~ g_{\rho\pi\pi}=6.08 \nonumber\\
&&f_\pi=0.0933\ \ {\rm GeV}.
\end{eqnarray}
Note that  for the $\pi'$ decay constant, $f_{\pi'}$, the value of 
0.02 GeV was used in Ref. [23]. As pointed out in Ref.[6], this value
might be overestimated because the mixing between the chiral pion field
and a massive pseudoscalar $q\bar{q}$ bound state should be considered.
Taking into account such mixing effects,  $f_{\pi'}$ was  re-estimated 
in the chiral Lagrangian framework and was found to be  $(1\sim 5) 
\times 10^{-3}$ GeV [6]. In our calculation, 
we use the most conservative value of $1\times 10^{-3}$ GeV  
and will comment on the effect of the larger $f_{\pi'}$ later.

This matrix element can be casted into the form [6]
\begin{eqnarray}
M=\sqrt{2}G_F(1+\chi)\left[\sum_\lambda L_{\sigma\lambda}H_\lambda
  +\left (1-\frac{m^2_{\pi^\prime}}{(m_u+m_d)m_\tau}
    \frac{\chi}{1+\chi}\right )L_{\sigma s}H_s\right],
\end{eqnarray}
where $L_{\sigma\lambda}$ and $L_{\sigma s}$ are the leptonic amplitudes, and
 $H_\lambda$  ($\lambda=0,\pm$) and $H_s$ are those involving hadrons. 
They are given by 
\begin{eqnarray}
L_{\sigma +}&=&0,\\
L_{\sigma 0}&=&\frac{m_\tau}{\sqrt{q^2}}\sqrt{m^2_\tau-q^2}
                 \delta_{\sigma +},\\
L_{\sigma -}&=&\sqrt{2}\sqrt{m^2_\tau-q^2}
                 \delta_{\sigma -},\\
L_{\sigma s}&=&\sqrt{m^2_\tau-q^2}\delta_{\sigma +},\\
H_\lambda &=& -\frac{2\sqrt{2}}{3}N\epsilon_\mu(q,\lambda)
            [(q_2-q_3)^\mu F_1(q^2,s_1)+(q_1-q_3)^\mu F_2(q^2,s_2)],\\
\label{Hs}
H_s &=& Nm_\tau C_{\pi^\prime}
              [s_1(s_2-s_3)F_3(q^2,s_1)+s_2(s_1-s_3)F_4(q^2,s_2)],
\end{eqnarray}
where $\epsilon(q,\lambda)$ is the polarization vector of the
virtual vector meson $a_1$. 

The amplitude of $\tau^+\rightarrow (3\pi)^+ \bar \nu_{\tau}$ 
can be obtained from that of 
$\tau^- \rightarrow (3\pi)^- \nu_{\tau}$ by the substitutions [6]:
$\chi \rightarrow \chi^*$, 
$L_{\sigma\lambda}\rightarrow (-1)^{\lambda}L_{-\sigma,-\lambda}$
and $L_{\sigma s}\rightarrow L_{-\sigma s}$. 

Following Ref. [6], we define two coordinate systems ($x,y,z$)
and ($x^*,y^*,z^*$) in the $(3\pi)^{\pm}$ rest frame.
Both systems have a common $y$-axis which is chosen along 
the $\vec{k}\times \vec{q}_3$ direction.
In the ($x,y,z$) system, the $z$-axis is along the direction
of $\vec{k}$, and the momentum $\vec{q}_3$ is in the (z,x)-plane with a
positive-$x$ component.  The  ($x^*,y^*,z^*$) system is related to 
the ($x,y,z$) system by 
a rotation by $\theta$ (the angle between $\vec{k}$ and $\vec{q_3}$)
with respect to the common $y$-axis, so that the $z^*$-axis is along 
$\vec{q}_3$. 
In terms of the five variables, $q^2$, $s_1$, $s_2$, $\theta$ and 
$\phi^*$, where $\phi^*$ is the azimuthal angle of $\vec{q}_1$
in the $(x^*,y^*,z^*)$ system,  we denote the differential decay rates 
of $\tau^- \rightarrow (3\pi)^- \nu_{\tau}$ 
and $\tau^+\rightarrow (3\pi)^+ \bar \nu_{\tau}$ by 
$\bar{G}(q^2,s_1,s_2,\cos\theta,\phi^*)$ and 
$G(q^2,s_1,s_2,\cos\theta,\phi^*)$, respectively.
Then a CP-conserving sum $\Sigma$ and a CP-violating difference $\Delta$
can be constructed [6]
\begin{eqnarray}
\Sigma &=& G(q^2,s_1,s_2,\cos\theta,\phi^*)
    +\bar{G}(q^2,s_1,s_2,\cos\theta,-\phi^*)\nonumber \\
       &=& \frac{G_F^2m_\tau}{2^{7}\pi^6}\frac{(1-q^2/m^2_\tau)^2}{q^2}
       |1+\chi|^2 \nonumber \\
& & \times \left\{ 2|H_-|^2 + \frac{m^2_\tau}{q^2}|H_0|^2
    +|1+\xi|^2|H_s|^2  - 2\frac{m_\tau}{\sqrt{q^2}}
         \left [ 1 + {\rm Re}(\xi)\right ]
         {\rm Re}(H_0 H^*_s) \right \},\\
\label{delta}
\Delta &=& G(q^2,s_1,s_2,\cos\theta,\phi^*)
    -\bar{G}(q^2,s_1,s_2,\cos\theta,-\phi^*) \nonumber \\
       &=& -2 \frac{G_F^2m_\tau}{2^{7}\pi^6}\frac{(1-q^2/m^2_\tau)^2}{q^2}
       |1+\chi|^2
\frac{m_\tau}{\sqrt{q^2}}
     {\rm Im}(\xi) {\rm Im}(H_0H^*_s),
\end{eqnarray}
where $\xi$ is given by
\begin{eqnarray}
\xi=-\frac{m^2_{\pi^\prime}}{(m_u+m_d)m_\tau}
    \left(\frac{\chi}{1+\chi}\right).
\end{eqnarray}

Choosing a weight function $w(q^2,s_1,s_2,\cos\theta,\phi^*)$,   
we can obtain a CP-violating observable $\langle w\Delta\rangle $ which
is obtained by integrating the quantity $w\Delta$ over the allowed phase space.
Following Ref. [6], we consider two types of CP-violating forward-backward 
asymmetries, $A_{1FB}$ and $A_{2FB}$, which are the $\nu_\tau$ 
($\bar{\nu}_\tau$) distribution with respect to the $\pi^+$ ($\pi^-$) 
direction in 
the $(3\pi)^-$ $\left ((3\pi)^+ \right )$ rest frame, with the 
respective weight functions ${\rm sign}[\cos\theta]$ and 
${\rm sign}[s_2-s_1] \cdot {\rm sign}[\cos\phi^*]$.
We also consider the optimal asymmetry, $A_{opt}$, which is defined with
the weight function $\Delta/\Sigma$. 
The statistical significance can be determined by
the quantity $\varepsilon=\langle w\Delta\rangle /\sqrt{\langle\Sigma\rangle
     \cdot\langle w^2\Sigma\rangle }$.
To observe this  CP-violating observable at the $2\sigma$ level, 
the required statistical significance should be 
$\varepsilon \ge 2/\sqrt {N_{\tau} {\rm Br}}$.  Hence the sensitivity to
probes of the coupling is proportional to $\sqrt{N_\tau}$.

Under the current constraint in (\ref{phiq}), 
i.e. $C^{(3)}_{\Phi L}/(\Lambda/{\rm TeV})^2 \le 1.4$, 
the number of $\tau$ required
to observe the effect of operator $O^{(3)}_{\Phi L}$ 
at the $2\sigma$ level is found to be (for $f_{\pi'}=1\times 10^{-3}$ GeV):
\begin{equation} \label{ntau}
N_{\tau}\ge \left \{ \begin{array}{ll}
                      7.2\times 10^6 ~~({\rm for}~ A_{1FB}) \\
                      0.5\times 10^6 ~~({\rm for}~ A_{2FB}) \\
                      0.8\times 10^5 ~~({\rm for}~ A_{opt}) 
                      \end{array} \right.
\end{equation}
So, at the tau-charm factory which will produce $1\times 10^7$ tau leptons per
year, it is possible to observe such CP-violating effects. If an effect is
not seen at the $2\sigma$ level, stronger constraints can be obtained for
the coupling strength of the operator under consideration,
\begin{equation}
\frac{\vert C^{(3)}_{\Phi L}\vert }{(\Lambda/{\rm TeV})^2} \le \left \{ \begin{array}{ll}
                    1.18 ~~({\rm for}~ A_{1FB}) \\
                    0.30 ~~({\rm for}~ A_{2FB}) \\
                    0.12 ~~({\rm for}~ A_{opt}) 
                      \end{array} \right.
\end{equation}

For the class B operators ($O_{LW}$, $O_{D\tau}$ and $O_{\tau W\Phi}$), we 
do not present their detailed analyses here. 
But we can roughly estimate the number of $\tau$ leptons required
to observe their effects from our results for the operator $O^{(3)}_{\Phi L}$.
As showed in Sec. 4, the upper bound on the coupling strength of a 
class B operator is $10^{-2}$ lower than that of $O^{(3)}_{\Phi L}$.
Since the number of $\tau$ leptons required
to observe the effects of an operator is proportional to $1/C_i^2$,
the number of $\tau$ leptons required
to observe the effects of a class B operator should be increased 
by a factor $10^4$ relative to that in (\ref{ntau}).  Hence,
under the current constraints, $10^9$ taus are needed in order to observe
the effects of a class B operator even with the most sensitive probe by 
$A_{opt}$.  So it is impossible to observe their effects in the
$3\pi$ mode of the tau decay at the 
tau-charm factory which is expected to produce $10^7$ taus per year.   

From Eqs.(\ref{Cpi}), (\ref{Hs}) and (\ref{delta}), we see that 
the CP-violating difference $\Delta$ is proportional to the
$\pi'$ decay constant $f_{\pi'}$ and thus the number of $\tau$ leptons 
required is proportional to $1/f_{\pi'}^2$. In our calculation we used
the most conservative value of $1\times 10^{-3}$ GeV. 
If we take the largest value of $5\times 10^{-3}$ GeV [6]
for $f_{\pi'}$, the number of $\tau$ leptons required to observe the
effects of the operators as given in Eq. (\ref{ntau}) will be lowered 
by a factor of 1/25. 

To conclude this section, let us discuss briefly the present experimental
situation.  The $\tau \rightarrow 3\pi\nu_\tau$ has been investigated by 
Argus [24] at Doris II and by OPAL [25], ALEPH [26] 
and, most recently, DELPHI [27] at LEP.  The statistics in these
experiments are several times to an order of magnitude smaller than that 
required by Eq. (\ref{ntau}).  Two models have been used by these
experiments to fit their data: One is the model by Isgur, Morningstar, 
and Reader (IMR) [22] and the other by Kuhn and Santamaria (KS) [22].  
The two models differ by their detailed parametrization [28] although 
both assume axial vector, i.e., $a_1$, dominance of the 3$\pi$ system.   
The model we used is close to that of IMR.  All data were found to be 
consistent with the two models although the detailed fits reveal some 
disagreement between the experimental data and the models.  For the 
OPAL collaboration which has made a detailed fit, both models were found
to overestimate the $\rho$-peak and underestimate the low value region of 
invariant mass of the $\pi^+\pi^-$ system.  In particular the IMR model 
was found to need a 14\% non-$a_1$ contribution which was represented as
a polynomial background.  In the most recent DELPHI analysis, the Dalitz 
plot and the invariant mass of the 3$\pi$ system were analyzed and 
found to be in reasonably good agreement with both the IMR and KS models.  
However, it is also found that the two models do not give a good fit for 
the $s_1$ and $s_2$ distributions for $s > 2.3$ GeV$^2$, where $s$ is the 
3$\pi$ invariant mass squared.  
\vspace{.5cm}

\begin{center} {\Large 6. Summary} \end{center}

We studied the dimension-six CP-violating $SU_L(2)\times U_Y(1)$ invariant 
operators involving the tau lepton, which could be generated by new physics 
at a higher energy scale. Under our criteria,
there are totally 10 such operators which are classified
into 4 classes. Since the class D operators, ($O_{\Phi L}^{ (1)}$, 
$O_{\Phi \tau}$ 
and $O_{\tau 1}$), only contribute to $H\tau\tau$ coupling, 
they are not constrained by current data and it
will be difficult to probe them at future colliders. 
 Class B operators ($O_{LW}$, $O_{D\tau}$ and $O_{\tau W\Phi}$) and Class C 
operators ($O_{LB}$, $O_{\tau B}$ and $O_{\tau B\Phi}$) give anomalous 
neutral currents and  are strongly constrained by
LEP experimental data. 
 Class A operator ($O_{\Phi L}^{ (3)}$) only contribute to charged current
and so far is only loosely constrained.
Although both Class A and Class B contribute to  charged current and
affect the tau hadronic decay, $\tau \rightarrow 3\pi \nu_{\tau}$,
only Class A operator will be possibly observable  at future 
tau-charm factory. The current strong limits on the Class B 
operators make their contributions to CP violation effect in 
$\tau \rightarrow 3\pi \nu_{\tau}$ unobservable.
 
To conclude, our analyses show that the CP-violating new physics, 
subject to existing experimental limits, are possibly observable 
in hadronic tau decays at the future tau-charm factory. The effective
operators (\ref{eq12}) - (\ref{eq21}) contain vertices of more 
complicated structures than the the 3-point vertices we have 
investigated in this article.  Their effects, in particular of 
operators given in Eqs. (\ref{eq12}), (\ref{eq13}), (\ref{eq14}), 
(\ref{eq16}) and (\ref{eq21}), will be investigated in future works.

Despite the fact that the decay $\tau \rightarrow 3\pi + \nu_\tau$ is an
ideal place to investigate the lepton CP-violation effect, the
validity of the model parametrization is not completely clear as 
discussed at the end of Sec. 5. In the future, a model independent 
parametrization of the process will be desirable.  Therefore, we 
see more theoretical effort is needed in order to reliably extract
any lepton CP-violation effect which may exist.

\vspace{.5cm}

\begin{center}{\Large  Acknowledgment}\end{center}

We thank Z. J. Tao  for discussions. J. M. Y. acknowledges JSPS for
the Postdoctoral Foreign Fellowship.
This work was supported in part by the U.S. Department of Energy, Division
of High Energy Physics, under Grant No. DE-FG02-94ER40817, by 
NSF of China and by the Grant-in-Aid for JSPS fellows from
the Japan Ministry of Education, Science, Sports and Culture. 
\vspace{1cm}

{\LARGE References}
\vspace{0.3in}
\begin{itemize}
\item[{\rm [1]}] K. Kobayashi, T. Maskawa, Prog. Theor. Phys.  {\bf 49}, 
                 652  (1973). 
\item[{\rm [2]}] For a review see A. G. Cohen, D. B. Kaplan and A. E. Nelson,
                  Annu. Rev. Nucl. Phys. {\bf  43}, 27  (1993). 
\item[{\rm [3]}] C. A. Nelson et al., Phys. Rev. D {\bf 50}, 4544 (1994).
\item[{\rm [4]}] M. Finkemeier and E. Mirkes, {\it Decay rates, structure 
                 functions and new physics effects in hadronic tau decays},
                 Proc. Workshop on Tau/Charm Factory (Argonne, 1995), 
                 ed. J. Repond, AIP Conf. Proc. No. 349 (New York, 1996) p.119.
\item[{\rm [5]}] U. Kilian et al., Z. Phys. C {\bf 62}, 413 (1994).
\item[{\rm [6]}]  S.Y. Choi, K. Hagiwara and  M. Tanabashi, 
                  Phys. Rev. D {\bf 52}, 1614 (1995).
\item[{\rm [7]}] Y. S. Tsai, hep-ph/9801274.
\item[{\rm [8]}] Y. S. Tsai,  Phys. Rev. D {\bf 51}, 3172 (1995). 
\item[{\rm [9]}] T. Huang, W. Lu and Z.J. Tao, Phys. Rev. D {\bf 55}, 
                 1643 (1997).
\item[{\rm [10]}] For a review, see G. Altarelli, R. Barbieri and 
                 F. Caravaglios, hep-ph/9712368.
\item[{\rm [11]}]  C. J. C. Burgess and H. J. Schnitzer, 
                 Nucl. Phys. B {\bf 228}, 454 (1983);
                 C. N. Leung, S. T. Love and S. Rao, Z. Phys. C {\bf 31}, 
                 433 (1986);
                 W. Buchmuller and D. Wyler, Nucl. Phys. B {\bf 268}, 
                 621 (1986). 
\item[{\rm [12]}]   K. Hagiwara, S. Ishihara, R. Szalapski and D. Zeppenfeld,
                 Phys. Lett. B {\bf 283}, 353 (1992);
                    Phys. Rev. D {\bf 48}, 2182 (1993). 
\item[{\rm [13]}]  G. J. Gounaris, F. M. Renard and C. Verzegnassi, 
                 Phys. Rev. D {\bf 52}, 451 (1995);
                  G. J. Gounaris, D. T. Papadamou and F. M. Renard,
                  hep-ph/9609437. 
\item[{\rm [14]}]  K. Whisnant, J. M. Yang, B.-L. Young and X. Zhang, 
                 Phys. Rev. D {\bf 56}, 467 (1997). 
\item[{\rm [15]}] J. M. Yang and  B.-L. Young, 
                  Phys. Rev. D {\bf 56}, 5907 (1997).
\item[{\rm [16]}] N. Wermes, {\it CP Tests and Dipole Moments in Tau-Pair
                                  Production Experiments}, 
                  in Proc. {\it Forth Workshop on Tau Lepton Physics}
                  (Estes Park, Colorado, 1996), ed. J. Smith, 
                  {\it Nucl. Phys. B (Proc. Suppl.)} to appear.
\item[{\rm [17]}] R. Escribano and E. Masso, Phys. Lett. B {\bf 395}, 
                  369 (1997).
\item[{\rm [18]}] S. M. Berman,  Phys. Rev. {\bf  112}, 267 (1958);
                  T. Kinoshita and A. Sirlin, Phys. Rev.  {\bf 113}, 1652  
                 (1959);
                  A. Sirlin, Rev. Mod. Phys.  {\bf 50}, 573  (1978);
                W. J. Marciano and A. Sirlin, Phys. Rev. Lett.  {\bf 61}, 
               1815 (1988).
\item[{\rm [19]}]  T. G. Rizzo, Phys. Rev. D {\bf 56}, 3074  (1997).
\item[{\rm [20]}]  M.-T. Dova, J. Swain and  L. Taylor, hep-ph/9712384.
\item[{\rm [21]}]  W. Li, talk at the {\it XVIII International Symposium
                   on Lepton Photon Interactions, Hamburg, July 28- August, 1
                   1997. }
\item[{\rm [22]}]  N. Isgur, C. Morningstar and C. Reader, 
                   Phys. Rev. D {\bf 39}, 1357 (1989);
                   J. H. Kuhn and A. Santamaria, Z. Phys. C {\bf 48}, 445 
                   (1990);
                   J. H. Kuhn and E. Mirkes, Phys. Lett. B {\bf 286}, 
                   381 (1992); Z. Phys. C  {\bf 56}, 661 (1992);
                   N. A. Tornqvist, Z. Phys. C {\bf 36}, 695 (1987);
                   M. G. Bower, Phys. Lett. B  {\bf 209}, 99 (1988);
                   Y. P. Ivanov, A. A. Osipov and M. K. Volkov,
                                 Z. Phys. C  {\bf 49}, 563 (1991);
                   R. Decker, E. Mirkes, R. Sauer and Z. Was, 
                            Z. Phys. C  {\bf 58}, 445 (1993); 
\item[{\rm [23]}] S. Jadach, J. H. Kuhn  and Z. Was, Comput. Phys. Commun.
                   {\bf  64},
                  275 (1991);
                  M. Jezabek, Z. Was, S. Jadach and J. H. Kuhn, {\it ibid.,}
                   {\bf 70}, 69 (1992);
                  S. Jadach, Z. Was, R. Decker  and J. H. Kuhn, {\it ibid.,}
                  {\bf 76}, 361 (1993).
\item[{\rm [24]}] ARGUS Collaboration,  Z. Phys. C {\bf 58}, 61 (1993).
\item[{\rm [25]}] OPAL  Collaboration,  Z. Phys. C {\bf 67}, 45 (1995);
                  {\it ibid.,} {\bf 75}, 593 (1997).
\item[{\rm [26]}] ALEPH Collaboration, Z. Phys. C {\bf 59}, 369 (1993).
\item[{\rm [27]}] DELPHI Collaboration, CERN-EP/98-14, Jan 30, 1998.
\item[{\rm [28]}] P. R. Poffenberg, Z. Phys. C {\bf 71}, 579 (1996),
                  identified the features in which the KS and IMR models
                  differ and found that the strong form factor has the
                  most influence on the fittings of the distribution 
                  shape and resonance parameters of the a$_1$. 
                  
\end{itemize}
\eject

Table 1 \\
The contribution status of dimension-six CP-violating operators 
to the tau couplings. 
The contribution of a CP-violating operator to a particular vertex is
marked by $\times$. 
\vspace{1cm}
\begin{center}

\large
\begin{tabular}{|l|c|c|c|c|}
\hline
 & & & & \\ 
 &$~W\nu \tau ~$ & $~Z\tau \tau~$ & $~\gamma \tau\tau~$ & $~H\tau\tau~$ \\ 
 & & & & \\ \hline

  $~~~O_{\Phi L}^{ (3)}$ &$\times $ & & &$\times$ \\ \hline
  $~~~O_{LW}$ & $\times $ & $\times $ &$\times $ & \\ \hline
  $~~~O_{D\tau}$     &$\times$ & $\times$ & &$\times $ \\ \hline
  $~~~O_{\tau W\Phi}$ &$\times$ & $\times$ & $\times$ & \\ \hline
  $~~~O_{LB}$ & & $\times $ & $\times $ & \\ \hline
  $~~~O_{\tau B}$ & & $\times $ & $\times $ & \\ \hline
  $~~~O_{\tau B\Phi}$ & &$\times$ &$\times $ & \\ \hline
  $~~~O_{\Phi L}^{ (1)}$ & & & &$\times$\\ \hline
  $~~~O_{\Phi \tau}$ & & & &$\times$ \\ \hline
  $~~~O_{\tau 1}$     & & & &$\times$ \\ \hline
\hline
\end{tabular}
\end{center}
\end{document}